# Relationship between blocking temperature and strength of interparticle interaction in magnetic nanoparticle systems

M. S. Seehra* and K. L. Pisane

Department of Physics & Astronomy, West Virginia University, Morgantown, WV 26506, USA

---

**Abstract**

In magnetic nanoparticle systems, the variation of the blocking temperature $T_B$ with the measuring frequency $f_m$ is often used to determine the strength of the interparticle interactions (IPI) through a parameter $\Phi$ or the Vogel-Fulcher temperature $T_o$. Presence of IPI is inferred if $T_o > 0$ and $\Phi = \Delta T_B / [T_B \Delta \log_{10} f_m] < 0.13$ where $\Delta$ signifies changes in $T_B$ and $f_m$. Here it is shown that these two parameters are related by the Eq. $\Phi = \Phi_o [1 - (T_o/T_B)]$ where $\Phi_o \approx 0.11$ to 0.15 is a constant of the system depending on the magnitudes of measuring frequency and the attempt frequency $f_o$ of the Néel relaxation. Experimental verification of this relationship is also presented using data on a variety of nanoparticle systems.

---

*Corresponding author. Email: mseehra@wvu.edu; Phone: 304-293-5098

## 1. Introduction

Magnetic properties of materials in reduced dimensions (thin films, wires and nanoparticles) continue to attract world-wide attention for two reasons: the emergence of new scientific phenomenon and potential applications of materials with reduced dimensions [1-4]. For magnetic nanoparticles (NPs), potential applications include tumor therapy by magnetic hyperthermia, targeted dug delivery, MRI contrast agents and biosensors [5-10]. An important property of magnetic NPs is their blocking temperature $T_B$ which separates the superparamagnetic (SPM) state for temperature $T > T_B$ from the frozen magnetic state for temperature $T < T_B$. For many biomedical applications, the NPs need to be in the SPM state at room temperature requiring $T_B < 300$ K so that the NPs are only magnetic in an applied magnetic field $H$ without any residual magnetization ($M$) when $H$ is removed. Thus determination of the blocking temperature of a nanoparticle system is usually the first property to be measured.

To understand the concept of the blocking temperature $T_B$, consider a nanoparticle of volume $V$ and anisotropy constant $K_a$. At a temperature $T$ and in the absence of any interparticle interactions (IPI), the rate of flipping $f$ of the magnetic moment of the particle against the energy barrier $E_a = K_a V$ is given by the Néel-Brown relaxation [11, 12]:

$$f = f_o \exp(-E_a/k_B T) \text{ --------- (1).}$$

In Eq. (1), $k_B$ is the Boltzmann constant and $f_o$ is the attempt frequency varying only weakly with temperature. An anisotropy related temperature $T_a = E_a/k_B$ may be conveniently defined. From Eq. (1), when $f$ becomes equal to $f_m$, the frequency of measurement, then system will appear to be blocked below the blocking temperature $T_B$ given by

$$T_B = T_a / \ln(f_o/f_m) \text{ --------- (2)}.$$

Computationally [13-16] and experimentally [17, 18], it is now established that the presence of interparticle interactions raises the magnitude of the blocking temperature $T_B$. In order to provide a measure of the IPI, Eq. (1) is often replaced by the Vogel-Fulcher law [19, 20] given by

$$f = f_o \exp[-E_a/k_B(T - T_o)] \text{ ---------- (3)}.$$

In Eq. (3), $T_o$ is an effective temperature representing the strength of the IPI among the NPs. Use of Eq. (3) leads to a new relation for $T_B$, replacing Eq. (2), in which $T_B$ is enhanced by $T_o$:

$$T_B = T_o + T_a / \ln(f_o/f_m) \text{ -------- (4)}.$$

Experimentally, the presence of IPI in a magnetic nanoparticle system can be detected by comparing the magnitudes of $T_B$ at a fixed $f_m$ with and without magnetic dilution since with magnetic dilution, such as coating of the particles with a surfactant, $T_B$ will be lower [17, 18] if IPI are present. An alternative approach to detecting the presence of IPI is determining how $T_B$ in a system varies with the change in the measuring frequency $f_m$. For this purpose, a quantity $\Phi$ has been defined as [20]:

$$\Phi = \Delta T_B / [T_B \Delta \log_{10} f_m] \text{ ----------- (5)}.$$

In Eq. 5, $\Delta T_B$ is the change in $T_B$ with change in $f_m$. Usually $T_B$ is defined as the position of the peak in the ac susceptibility $\chi''$ measured at a particular $f_m$. The magnitude of $\Phi$ is known to vary with the strength of IPI; $\Phi \geq 0.13$ for non-interacting particles, $0.05 < \Phi < 0.13$ for interacting particles with $\Phi$ decreasing with increase in the strength of IPI and $\Phi < 0.05$ for spin-glasses [20]. Since both $T_o$ and $\Phi$ have been used in the literature to provide a measure of the strength of the IPI in a system, it is instructive to determine whether a relationship exists between these two parameters. In this paper, we derive such a relationship and verify its validity using our own yet unpublished data on maghemite NPs and data from published papers in a number of nanoparticle systems where the magnitude of both $T_o$ and $\Phi$ are available.

## 2. Derivation of the relationship between $T_o$ and $\Phi$

Consider measurements of $T_B$ in a system using two different measuring frequencies $f_m(1)$ and $f_m(2)$ with $f_m(2) > f_m(1)$. Using Eq. (4) yields

$$T_B(1) = T_o + T_a/[\ln f_o - \ln f_m(1)] \text{------- (6)}$$

$$T_B(2) = T_o + T_a/[\ln f_o - \ln f_m(2)] \text{------- (7).}$$

Using equations (6) and (7), $\Delta T_B = T_B(2) - T_B(1)$ is calculated and substituted in Eq. (5) to determine $\Phi$. After some simplifications, the following equations are derived without any approximation:

$$\Phi = \Phi_o\{1 - [T_o/T_B(1)]\} \text{------ (8)}$$

$$\Phi_o = 2.3026/\{\ln[f_o/f_m(2)]\} \text{------------ (9).}$$

From Eq. (9), the magnitude of $\Phi_o$ is a constant of the system in that it depends on the attempt frequency $f_o$ and measuring frequency $f_m(2)$. Usually, $f_o$ varies between $10^9$ Hz to $10^{12}$ Hz for different systems and $f_m$ in commercially available experimental systems can be varied from a low value of 0.1 Hz to high value of $10^4$ Hz. As examples using Eq. (9), if $f_o = 10^{10}$ ($10^9$) Hz and $f_m(2) = 10^3$ Hz, then $\Phi_o = 0.143$ (0.167) is obtained. Similarly for $f_o = 10^{12}$ Hz and $f_m(2) = 10^4$ Hz, $\Phi_o = 0.125$ is obtained. For the often quoted $\Phi_o = 0.13$, $f_o/f_m(2) = 4.9 \times 10^7$ is needed.

Eq. (8), relating $\Phi$ to the fractional change $T_o/T_B(1)$, along with Eq. (9) for $\Phi_o$ and the experimental verification of Eq. (8) given below are the new results of this paper.

## 3. Experimental Verification

For experimental verification of Eq. (8) connecting $\Phi$, $T_o$ and $\Phi_o$, the literature was searched to find magnetic nanoparticle systems where magnitudes of $\Phi$, $T_B(1)$, and $T_o$ have been published by the authors. The plot of $\Phi$ versus $T_o/T_B(1)$ should yield a straight line if Eq. (8) is valid with the intercept yielding the magnitude of $\Phi_o$. Such a plot is shown in Fig. 1 with each data point belonging to a different system. Overall, the data follows the predicted linear variation of $\Phi$ with $T_o/T_B(1)$ with the two dotted lines drawn for $\Phi_o = 0.11$ and $\Phi_o = 0.15$ bracketing most of the data points. The scatter in the data points around the predicted linear behavior likely results from different magnitudes of the ratio $f_o/f_m(2)$ and hence $\Phi_o$ in different systems. In Fig. 1, the numbers in [ ] adjacent to each system represent reference to the publication from which the data is taken.

## 4. Conclusions

In this paper, it is shown that the two parameters $\Phi$ and $T_o$ usually used to determine the strength of the interparticle interactions in magnetic nanoparticle systems are related by Eq. (8) with $\Phi_o$ given by Eq. (9). The experimental verification of this relation is presented in Fig. 1 using data on a variety of nanoparticle systems. The magnitude of $\Phi_o \approx 0.13$ often quoted in papers really depends on the ratio $f_o/f_m(2)$ for each system and so it may differ somewhat from this magnitude as shown in representative calculations given above. The above derivation depends on the validity of the Vogel-Fulcher relation (Eq. 3) for a system.

**Acknowledgement**

This work was supported in part by a grant from the U.S. National Science Foundation, Grant # DGE-1144676.

**Figure Captions:**

Figure 1: Verification of Eq. (8) by plotting $\Phi$ versus $(T_o/T_B)$ for different systems listed in the legend of the figure. The solid circles are experimental data points with experimental uncertainties shown for cases where this information was available. The two dotted lines represent Eq. (8) with $\Phi_o = 0.15$ and $0.11$. The numbers in [ ] give reference to the paper from which data are taken.

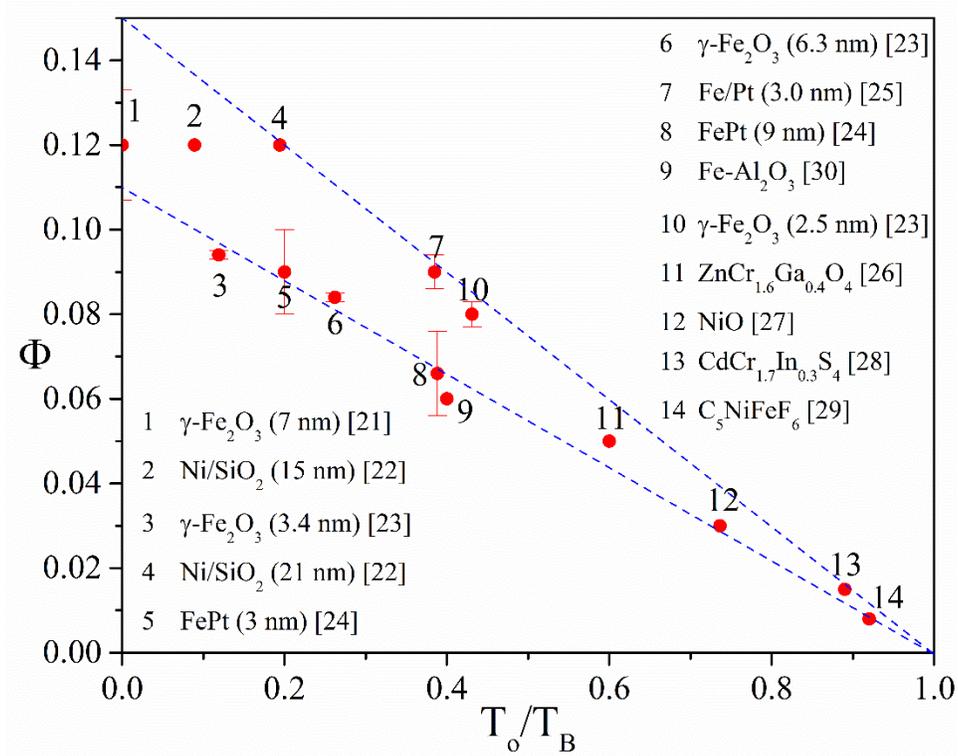

Figure 1: Verification of Eq. (8) by plotting $\Phi$ versus $(T_o/T_B)$ for different systems listed in the legend of the figure. The solid circles are experimental data points with experimental uncertainties shown for cases where this information was available. The two dotted lines represent Eq. (8) with $\Phi_o = 0.15$ and $0.11$. The numbers in [ ] give reference to the paper from which data are taken.